\documentclass[prb,showpacs,floatfix,superscriptaddress,twocolumn]{revtex4}
\usepackage{amsmath}
\usepackage{graphicx}
\usepackage{float}
\usepackage{bm}
\usepackage{color}
\begin{document}

\bibliographystyle{apsrev}

\title{Microwave conductivity in the ferropnictides with specific application
to Ba$_{1-x}$K$_x$Fe$_2$As$_2$
}
\author{E. Schachinger}
\email{schachinger@itp.tu-graz.ac.at}
\affiliation{Institute of Theoretical and Computational Physics,
Graz University of Technology, A-8010 Graz, Austria}
\author{J. P. Carbotte}
\affiliation{Department of Physics and Astronomy, McMaster University, Hamilton,
Ontario, Canada N1G 2W1}
\affiliation{The Canadian Institute for Advanced Research, Toronto, Ontario,
  Canada M5G 1Z8}
\date{\today}
\begin{abstract}
We calculate the microwave conductivity of a two band superconductor
with $s^\pm$ gap symmetry. Inelastic scattering is included approximately
in a BCS model augmented by a temperature dependent
quasiparticle scattering rate assumed, however, to be frequency independent.
The possibility that the $s$-wave gap on
one or the other of the electron or hole pockets is anisotropic is
explored including cases with and without gap nodes on the Fermi surface.
A comparison of our BCS results with those obtained in the
Two Fluid Model (TFM)  is provided as well as with the case of the
cuprates where the gap has $d$-wave symmetry and with experimental
results in Ba$_{1-x}$K$_x$Fe$_2$As$_2$. The presently available microwave
conductivity data in this material provides strong evidence for large
anisotropies in the electron pocket $s$-wave gap. While a best fit
favors a gap with nodes on the Fermi surface this disagrees with some
but not all penetration depth measurements which would favor a
node-less gap as do also thermal conductivity and nuclear magnetic
resonance data.
\end{abstract}
\pacs{74.25.Nf, 74.20.Rp, 74.25.Fy 74.70.-b}
\maketitle
%
%

\section{Introduction}

In a dirty isotropic $s$-wave BCS superconductor a so-called
`coherence peak' appears in the microwave conductivity at some
reduced temperature $t = T/T_c$ slightly below one. Here $T_c$ is
the critical temperature at which the material becomes superconducting.
Reducing the residual impurity scattering rate pushes the peak closer
to $T=T_c$ and reduces its amplitude. In the clean limit no coherence
peak remains. Similar trends also characterize the microwave response
as the probing frequency is increased. While the weak coupling limit
of Eliashberg theory reproduces the BCS results described above, an increase
in inelastic scattering provides additional damping effects which
decrease the coherence peaks of BCS theory but, provided they are not
too large, these remain.

A different behavior was observed in cuprate superconductors very
early on and this has been confirmed in later experiments. There is
no coherence peak just below $T_c$, rather a peak which can have an
even larger amplitude is seen at small values of the reduced temperature
much below $T_c$. The peak is sensitive to residual scattering. For
example doping with small amounts of Zn or Ni can greatly reduce the
peak height and can also shift the temperature $T$ at which it
occurs.\cite{bonn94} One can understand these observations semiquantitatively
within the phenomenological two fluid model (TFM). Only the normal
fluid component, $n_N(T)$, enters the real (absorptive) part of the
conductivity, $\sigma_1(T,\omega)$. As the temperature is reduced
towards zero more and more of the charge carriers enter the condensate
leaving less and less normal fluid and so decreases $\sigma_1(T,\omega)$.
But at the same time the quasiparticle scattering time $\tau(T)$
increases and this increase can be sufficiently fast so as to
more than compensate in $\sigma_1(T,\omega)$ for the drop in
$n_N(T)$ resulting in its net increase with decreasing $T$. But
$\tau(T)$ can not increase indefinitely with $T\to 0$. Eventually
it hits a maximum value set by the finite, non-zero impurity
scattering time even for the cleanest of samples. When this limit
is reached, the microwave conductivity can no longer increase and, thus,
will start to drop tracking the reduction in $n_N(T)$ as $T\to 0$.
For a $d$-wave superconductor $n_N(T)$ is linear in $T$ for
small $T$ but for isotropic $s$-wave it becomes exponentially small
for $T$ less than the gap amplitude $\Delta_s(T)$. This clearly
moves the temperature at which the drop in $\sigma_1(T,\omega)$ is
to be expected to higher values of $T$ and should also make the drop
more precipitous in isotropic $s$-wave than it is for $d$-wave.

In Sec.~\ref{sec:2} we begin with a review of the application of the
two fluid model to understand the microwave data observed in the
cuprates due to the collapse of the inelastic scattering at low
temperatures. We also provide a summary of the good fit to data obtained
within generalized Eliashberg theory for $d$-wave symmetry and low
energy cutoff applied to the electron-boson spectral density. This
cutoff is due to the opening of a spin gap and is identified as the
microscopic origin of reduced inelastic scattering. Next we proceed
to show that the data can also be understood reasonably well within
a simpler BCS approach but with a phenomenological temperature
dependent but frequency independent scattering rate
$\tau^{-1}_\textrm{BCS}(T)$. These results alow us to understand better
the limitations as well as the strength of the TFM approach.
Having established the usefulness of the BCS formulation, the method
is extended in Sec.~\ref{sec:3} to the case of an $s$-wave superconductor
including two bands (electron and hole pockets) each with
a different gap value. The gaps could be isotropic with opposite sign
corresponding to $s^\pm$-symmetry or one of the gaps could have nodes
on the Fermi surface. The specific case of Ba$_{1-x}$K$_x$Fe$_2$As$_2$
(FeAs-122) is treated extensively and comparison with experiment is made.
Conclusions and a summary are found in Sec.~\ref{sec:4}

\section{Microwave conductivity , penetration depth, and scattering rates}
\label{sec:2}

Many formulations of the optical conductivity start from a Kubo formula for
the current-current correlation function. Some use a finite temperature
Matsubara formalism with final analytic continuation to real frequencies
done with Pad\'e approximants.\cite{nicol91}
Others proceed within a real frequency axis
formalism\cite{carb95,mars96,schach98} for $s$ or $d$-wave gap symmetry.
For infinite
free electron bands the integral over the energy can be performed
analytically and the remaining integral done numerically. The input
are the solutions of the appropriate Eliashberg equations which follow once
the electron-boson spectral density $\alpha^2F(\omega)$ is specified.
For an electron-phonon system $\alpha^2F(\omega)$ would describe the
phonon exchange while for coupling dominantly to spin fluctuations 
as is envisaged in the nearly antiferromagnetic Fermi-liquid
model\cite{millis90}
(NAFLM) of the cuprates it describes the exchange of over-damped
spin waves. We refer the reader to some of this vast literature and,
here, we will not give details.\cite{schach03}

The microwave conductivity of a superconductor is calculated from
\begin{widetext}
\begin{equation}
  \label{eq:1}
  \sigma_1(T,\nu) = \frac{e^2\pi}{\nu}\sum\limits_{\textbf{k}}
2 v^2_{\textbf{k}x}\int\limits_{-\infty}^\infty\!d\omega\,
\left[ f(T,\nu+\omega)-f(T,\omega)\right]
\left[ A(\textbf{k},\omega)
A(\textbf{k},\nu+\omega)+B(\textbf{k},\omega)B(\textbf{k},\nu+\omega)
\right]. 
\end{equation}
Here $e$ is the charge on the electron, $\nu$ the microwave frequency,
$f(T,\omega)$ the Fermi-Dirac thermal occupation factor at temperature
$T$ and $A(\textbf{k},\omega)$ and $B(\textbf{k},\omega)$ are, respectively,
the usual charge carrier spectral density and the Gor'kov anomalous
equivalent which is zero in the normal state. Finally, $v_{\textbf{k}x}$
is the $x$ component of the electron velocity at momentum \textbf{k}.
The formula for the London penetration
depth which is related to the zero frequency limit of the imaginary part
of the optical conductivity by
\begin{equation}
  \label{eq:2}
  \frac{1}{\lambda^2_L} = \lim_{\nu\to 0}\frac{4\pi\nu}{c^2}\sigma_2(T,\nu),
\end{equation}
with $c$ the velocity of light is determined by
\begin{equation}
  \label{eq:3}
  \frac{1}{\lambda^2_L} = \frac{4\pi e^2}{c^2}\lim_{\nu\to 0}
  \sum\limits_\textbf{k}
  2 v^2_{\textbf{k}x}\int\limits_{-\infty}^\infty\!d\omega'
  \int\limits_{-\infty}^{\infty}\!d\omega''\,\frac{f(T,\omega'')-
  f(T,\omega')}{\omega''-\omega'}
  \lim_{\textbf{q}\to 0}2B(\textbf{k}+\textbf{q},\omega')
    \,B(\textbf{k},\omega'').
\end{equation}
\end{widetext}
The spectral densities $A(\textbf{k},\omega)$ and $B(\textbf{k},\omega)$
are obtained in the standard way from the $2\times 2$ Nambu matrix
Green's function ${\cal G}(\textbf{k},i\omega_n)$ with $i\omega_n$
the imaginary Matsubara frequencies, $i\omega_n = i\pi T(2n+1),\, n = 0,\pm 1,
\pm2, \dots$ In an Eliashberg formulation of the theory of
superconductivity inelastic scattering is included for an electron-boson
exchange mechanism. The superconducting gap function acquires a frequency
dependence and the bare frequency $\omega$ is renormalized to
$\tilde{\omega}(\omega) = \omega- \Sigma(\omega)$ with $\Sigma(\omega)$ the
charge carrier self energy. The static limit of the Eliashberg theory
reduces to BCS theory which deals only with impurity scattering which is
static and no retardation is included in the pairing potential.

As we are going to present results based only on BCS theory throughout
the paper we need to discuss the formalism applied here. It is based on the
mixed symmetry model by Sch\"urrer {\it et al.}\cite{schuerrer98}
and starts with a mixed symmetry order parameter
$\Delta_{sd}(\theta) = \Delta_s+\Delta_d\sqrt{2}\cos(2\theta)$
defining an $s+d$ symmetric order parameter. Here, $\Delta_s$ is the
$s$-wave symmetric component and $\Delta_d$ is the amplitude of the
$d$-wave component. $\theta$ is the polar angle on the cylinder
symmetric Fermi surface. We introduce, furthermore, the anisotropy
parameter $\alpha$ by
\begin{equation}
  \label{eq:alpha}
  \Delta_s = \alpha\Delta_0,\quad \Delta_d = \sqrt{1-\alpha^2}\Delta_0
\end{equation}
which ensures that
\begin{equation}
  \label{eq:alpha1}
  \Delta_0 = \sqrt{\left\langle\Delta_{sd}^2(\theta)\right\rangle_\theta},
\end{equation}
with $\langle\dots\rangle_\theta$ the Fermi surface average. The gap
$\Delta_0$ is, furthermore, assumed to have the standard BCS temperature
dependence. According to Eq.~\eqref{eq:alpha} $\alpha=0$ gives the
pure $d$-wave symmetric case, while $\alpha=1$ corresponds to the
isotropic $s$-wave case. The corresponding BCS equations for the
renormalized frequencies $\tilde{\omega}(\omega)$ and renormalized
gaps $\tilde{\Delta}_{sd}(\omega)$ at one temperature $T<T_c$ are
then given in a real axis notation by
\begin{subequations}
\label{eq:BCS}
\begin{eqnarray}
  \label{eq:BCSa}
  \tilde{\omega}(\omega) &=& \omega + i\tau^{-1}_\textrm{BCS}
  \left\langle\frac{\tilde{\omega}(\omega)}{\sqrt{\tilde{\omega}^2
  (\omega)-\tilde{\Delta}^2_{sd}(\omega,\theta)}}\right\rangle_\theta\\
  \label{eq:BCSb}
  \tilde{\Delta}_s(\omega) &=& \Delta_s+i\tau^{-1}_\textrm{BCS}
  \left\langle\frac{\tilde{\Delta}_{sd}(\omega,\theta)}
   {\sqrt{\tilde{\omega}^2
  (\omega)-\tilde{\Delta}^2_{sd}(\omega,\theta)}}\right\rangle_\theta\\
  \label{eq:BCSc}
  \tilde{\Delta}_d(\omega) &=& \Delta_d\\
  \label{eq:BCSd}
  \tilde{\Delta}_{sd}(\omega) &=& \tilde{\Delta}_s(\omega)+\tilde{\Delta}_d
  (\omega)\sqrt{2}\cos(2\theta).
\end{eqnarray}
\end{subequations}
Here, $\tau^{-1}_\textrm{BCS}$ is the elastic quasiparticle (QP)
impurity scattering rate which is temperature and frequency independent.
For convenience, we introduce
\begin{equation}
  \label{eq:x}
  x = \frac{\alpha}{\alpha+\sqrt{1-\alpha^2}},
\end{equation}
which, multiplied by 100, gives the percentage of the $s$-wave gap
$\Delta_s$ contained in $\Delta_{sd}(\theta)$. This gap will have nodes
on the Fermi surface as long as $\alpha\le\sqrt{2/3}$ ($x \le 0.59$)
in a clean limit system. Nevertheless, because of Eq.~\eqref{eq:BCSb}
the nodes on the Fermi surface can be lifted even for $\alpha <
\sqrt{2/3}$ if the residual resistivity of the sample is large enough.
This has also been discussed recently by Mishra {\it et al.}\cite{mishra09}

The complex optical conductivity $\sigma(T,\nu)$ at temperature $T$ and
frequency $\nu$ is calculated from the Kubo formula
\begin{widetext}
\begin{equation}
  \label{eq:sigma}
  \sigma(T,\nu) = \frac{\Omega^2_p}{4\pi}\frac{i}{\nu}\left\langle
  \int\limits_0^\infty\!d\omega\,\textrm{tanh}\left(\frac{\beta\omega}{2}
  \right)
  \left[J(\omega,\nu)-J(-\omega,\nu)\right]\right\rangle_\theta,
\end{equation}
with $\Omega_p$ the plasma frequency, $\beta = 1/(k_BT)$, and
\begin{equation*}
  2J(\omega,\nu) = \frac{1-N(\omega,\theta)N(\omega+\nu,\theta)
  -P(\omega,\theta)P(\omega+\nu,\theta)}{E(\omega,\theta)+
  E(\omega+\nu,\theta)}
  +\frac{1+N^\star(\omega,\theta)N(\omega+\nu,\theta)+
    P^\star(\omega,\theta)N(\omega+\nu,\theta)}{E^\star(\omega,\theta)-
    E(\omega+\nu,\theta)},
\end{equation*}
\end{widetext}
where $\star$ indicates the complex conjugate.
Here, $E(\omega,\theta) = \sqrt{\tilde{\omega}^2(\omega+0^+)-
\tilde{\Delta}^2_{sd}(\omega+i0^+,\theta)}$,
$N(\omega,\theta) = \tilde{\omega}(\omega+i0^+)/E(\omega,\theta)$, and
$P(\omega,\theta) = \tilde{\Delta}_{sd}(\omega+i0^+,\theta)/E(\omega,\theta)$.
The London penetration depth can then be calculated using Eq.~\eqref{eq:2}
or, as was demonstrated by Modre {\it et al.},\cite{modre98} more
conveniently in an imaginary axis representation of Eqs.~\eqref{eq:BCS}

It is important to stress before going on to a discussion of BCS results
as well as results based on the TFM which is favored in the analysis
of data provided in experimental papers\cite{bonn94,hashimoto09}
that Eliashberg theory
provides a good understanding of both microwave conductivity and
penetration depth in terms of a $d$-wave symmetry gap function and an
electron-boson spectral density which describes the
coupling to over-damped spin fluctuations with a low frequency cutoff.
This provides a temperature dependent inelastic QP scattering rate
which can get small at low temperatures where it is limited only by the
residual elastic impurity scattering. But the inelastic
QP scattering is also
unavoidably frequency dependent and
temperature and frequency dependence are
related to each other. By contrast, in BCS theory the QP scattering rate
$\tau^{-1}_{\textrm{BCS}}$ is frequency independent as it is also the
case for the TFM which gives as a result the temperature dependent
scattering rate $\tau^{-1}_\textrm{TFM}(T)$. In the simplest
case (no vertex corrections) the optical scattering rate in the
normal state is just twice the QP scattering rate.
 Nevertheless, one can model the inelastic QP scattering
through a phenomenological temperature dependent
$\tau^{-1}_{\textrm{BCS}}(T)$ but this cannot be exact as we will
elaborate upon later. In the TFM, on the other hand,
one assumes that the superfluid density
at temperature $T$, $n_S(T)$, plus the normal fluid density $n_N(T)$
add up to the total electron density per unit volume in the normal state,
denoted $n$. The London penetration depth
$1/\lambda^2_L(T) = \frac{4\pi e^2}{c^2}\frac{n_s(T)}{m}$ where $m$
is the electron mass. One can get the normal fluid density
from
\begin{equation}
  \label{eq:5}
  \frac{e^2n_N(T)}{m} = \frac{c^2}{4\pi}\frac{1}{\lambda^2_L(0)}\left[1-
  \frac{\lambda^2_L(0)}{\lambda^2_L(T)}\right].
\end{equation}
An optical scattering rate which we denote with
$\tau^{-1}_{\textrm{TFM}}(T)$ can then be defined in terms of the
microwave conductivity as
\begin{equation}
  \label{eq:6}
  \tau^{-1}_\textrm{TFM}(T) = \frac{c^2}{4\pi}\frac{1}{\lambda^2_L(0)}
  \frac{1-\lambda^2_L(0)/\lambda^2_L(T)}
  {\sigma_1(T)}
\end{equation}

\begin{figure}[tp]
  \includegraphics[width=9cm]{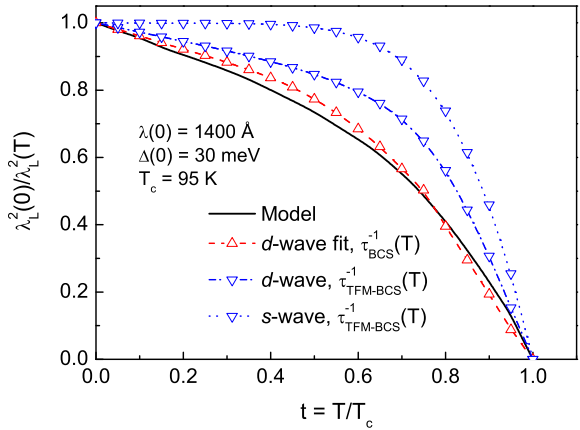}
  \caption{(Color online)
The normalized inverse square of the penetration depth
$\lambda^2_L(0)/\lambda^2_L(T)$ vs reduced temperature $t$. The solid
(black) curve are results from an Eliashberg calculation
(Ref.~\onlinecite{schach97}) which fit well
experimental results (Ref.~\onlinecite{bonn93}) on an
optimally doped YBCO single crystal. The open (red) up-triangles connected by
a dashed line are the results of our BCS $d$-wave fit based on the
scattering rate $\tau^{-1}_{\textrm{BCS}}(T)$.
The open (blue) down-triangles connected by a dashed-dotted line
are the result of a BCS $d$-wave calculation using the scattering rate
$\tau^{-1}_{\textrm{TFM-BCS}}(T)$ derived from the two fluid model
while the open (blue) down-triangles connected by a dotted line
give the corresponding result of a BCS $s$-wave calculation.
}
  \label{fig:2}
\end{figure}

Before dealing with the ferropnictide superconductors it will prove
useful to start with a very brief review of the situation in the
cuprates. These are $d$-wave superconductors but the usual isotropic
$s$-wave Eliashberg equations can easily be generalized to include
a momentum dependent superconducting gap which in two dimensions
can be taken to vary as $\Delta_0\cos(2\theta)$ with $\theta$ an
angle on the circular Fermi surface in the CuO$_2$ Brillouin zone.
Details can be found in our previous papers\cite{carb95,mars96,schach98}
where we considered data for the penetration depth\cite{schach97} and the
microwave conductivity\cite{schach98} in optimally doped
YBa$_2$Cu$_3$O$_{6.95}$ (YBCO) single crystals and find that an excellent
fit to both sets of data can be obtained with a spin fluctuation
spectral form\cite{millis90} (MMP form)
\begin{equation}
  \label{eq:4}
  \alpha^2F(\omega) = I^2\frac{\omega/\omega_{SF}}{1+(\omega/\omega_{SF})^2},
\end{equation}
where $I$ is the electron-spin fluctuation coupling strength and
$\omega_{SF}$ a characteristic spin fluctuation energy taken to be
$30\,$meV. To fit the microwave data a low frequency cutoff of
$\omega_c(T=0) = 2.1\,T_c$ was applied on the MMP form of Eq.~\eqref{eq:4}.
In the NAFLM\cite{millis90} this cutoff can be thought of as arising
from the formation of a spin gap in the superconducting state.
This concept was already introduced by  Nuss {\it et al.}\cite{nuss91}
and Nicol and Carbotte\cite{nicol91a} within the Marginal Fermi Liquid
Model (MFLM) of
Varma {\it et al.}\cite{varma89} to account for the gaping of the
spin and charge susceptibility brought about by the condensation
into Cooper pairs. The frequency cutoff
$\omega_c(T)$ was taken to decrease with increasing $T$ according to a BCS mean
field temperature dependence. The resulting normalized microwave
conductivity $\sigma_1(T)/\sigma_1(T_c)$ is displayed as a function of
the reduced temperature $t = T/T_c$ as the solid (black) curve in
%
%
\begin{figure}[tp]
  \vspace{-0.5cm}
  \includegraphics[width=9cm]{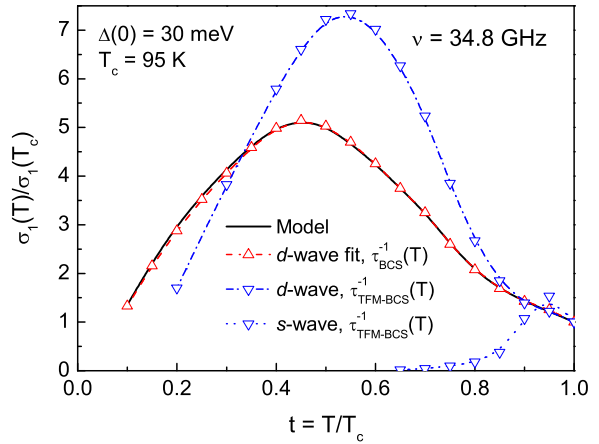}
  \caption{(Color online)
Normalized microwave conductivity $\sigma_1(T)/\sigma_1(T_c)$
vs reduced temperature $t$. The solid (black) line are the results from
an Eliashberg calculation of Ref.~\onlinecite{schach98} which fit
well experimental results (Ref.~\onlinecite{bonn94})
for a twin-free optimally doped YBCO single crystal. The
(red) open up-triangles represent our fit to the solid (black) curve
using BCS $d$-wave with a temperature dependent scattering rate
$\tau^{-1}_{\textrm{BCS}}(T)$. The open (blue) down-triangles connected by a
dashed-dotted line are additional BCS $d$-wave results obtained now
with the scattering rate $\tau^{-1}_{\textrm{TFM-BCS}}(T)$ derived from the
two fluid model. Finally, the open (blue) down-triangles connected by the
dotted line give the results of a BCS $s$-wave theory calculation using
the scattering rate $\tau^{-1}_{\textrm{TFM-BCS}}(T)$.
}
  \label{fig:1}
\end{figure}
Fig.~\ref{fig:1} for the microwave frequency $34.8\,$GHz used in
experiments.\cite{bonn94} The fit to the data is not shown here but
it was very good. When the same model is applied to the London
penetration depth an equally good fit to the data reported by
D. A. Bonn {\it et al.}\cite{bonn93} was obtained and it is shown
as the solid (black) line in Fig.~\ref{fig:2} for the normalized square
%
%
of the penetration depth $\lambda^2_L(0)/\lambda^2_L(T)$ in optimally
doped YBCO single crystals as a function of the reduced temperature $t$.
The model also provides a good fit to corresponding thermal conductivity
data.\cite{schach98a}

In this paper we will use the solid (black) curves
of Figs.~\ref{fig:1} and \ref{fig:2} for microwave conductivity and
penetration depth as representative of the cuprates and investigate whether
or not a simpler formulation of microscopic theory, namely BCS theory,
with a phenomenological temperature dependent scattering rate
$\tau^{-1}_\textrm{BCS}(T)$ can also provide a good understanding
of the data.

We begin with a fit to the microwave conductivity data of Fig.~\ref{fig:1}
[solid (black) line]. The open (red) up-triangles are our fit with the
%
%
\begin{figure}[tp]
  \includegraphics[width=9cm]{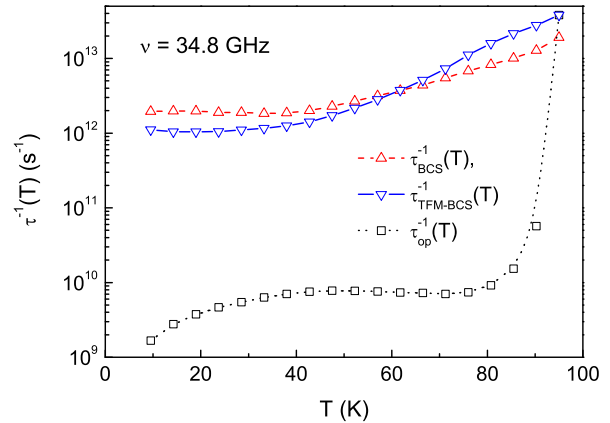}
  \caption{(Color online)
Scattering rates $\tau^{-1}$ in inverse seconds as a function
of temperature $T$
for an optimally doped YBCO single crystal.
The scattering rate $\tau^{-1}_{\textrm{BCS}}(T)$ [open (red) up-triangles]
was obtained by a BCS $d$-wave fit to the microwave
conductivity [solid (black) line in Fig.~\ref{fig:1}] and Born impurity
scattering. Finally, the (black) open squares present the optical
scattering rate $\tau^{-1}_\textrm{op}(T)$, Eq.~\eqref{eq:Drude},
at the microwave
frequency $\nu$. Note, that in the normal state at $T=95\,$K the
optical scattering rate is precisely twice the QP scattering rate
$\tau^{-1}_\textrm{BCS}(T=95\,\textrm{K})$. The open (blue) down-triangles,
finally, are the results of Eq.~\eqref{eq:TFMnew} as described in the text.
}
  \label{fig:3}
\end{figure}
the corresponding quasiparticle scattering rate denoted
$\tau^{-1}_\textrm{BCS}(T)$ shown as the open (red) up-triangles in
Fig.~\ref{fig:3}. The fit is not unique and corresponds to a choice
of least residual scattering at $T=0\,$, consistent with the
normalized data of Fig.~\ref{fig:1}. This choice is partially motivated
by the recognized fact that the cuprates are known to be rather pure.
Other fits all would have larger values of $\tau^{-1}_\textrm{BCS}(T)$
at $T=T_c$ as well as residual impurity scattering at $T=0$. This
ambiguity disappears if the plasma frequency $\Omega_p$ of
Eq.~\eqref{eq:sigma} is known. As a first check on the validity of
our phenomenological $\tau^{-1}_\textrm{BCS}(T)$ we can use it to
calculate other properties. In Fig.~\ref{fig:2} the (red) open
up-triangles represent our results for the normalized inverse square
of the penetration depth $\lambda^2_L(0)/\lambda^2_L(T)$ vs $T$. We see good,
although not perfect agreement with the solid (black) curve. This
demonstrates that a BCS approach with phenomenological
$\tau^{-1}_\textrm{BCS}(T)$ fits to the microwave conductivity data
cannot represent the penetration depth data which goes with it quite
as accurately as we can with Eliashberg theory. Nevertheless, the
fit is quite good and shows clearly that the simpler BCS approach
used here can be applied with confidence to other systems such as
the ferropnictides.

Moreover, one can define a TFM scattering rate based on our BCS calculation
without reference to the plasma frequency $\Omega_p$. We define
 \begin{equation}
   \label{eq:TFMnew}
   \tau^{-1}_\textrm{TFM-BCS}(T) = \frac{1-\lambda^2_L(0)/
   \lambda^2_L(T)}{\sigma'_1(T)},
 \end{equation}
where $\sigma'_1(T)$ is in computer units defined without the factor
$\Omega_p^2/(4\pi)$ in front of the right hand side of Eq.~\eqref{eq:sigma}.
Results are shown in Fig.~\ref{fig:3} as the open (blue) down-triangles.
The points cross the open (red) up-triangles for the quasiparticle
scattering rate $\tau^{-1}_\textrm{BCS}(T)$ of BCS theory and show that
the two scattering rates are not the same, they have a different
temperature dependence. We might have expected them to differ only
by a constant factor of two which is the relation expected to hold
between optical and QP residual scattering rates. But this doesn't
hold here and shows the limitation of the concept of a TFM based
scattering rate. We can even go further and use
$\tau^{-1}_\textrm{TFM-BCS}(T)/2$ as an effective temperature dependent
QP scattering rate in new BCS calculations. When this is done, we get
the open (blue) down-triangles connected by a dashed-dotted line in
Figs.~\ref{fig:1} and \ref{fig:2} for the microwave conductivity
and the penetration depth, respectively. The agreement with our
model data [solid (black) line] is very poor. In particular, the peak
in $\sigma_1(T)/\sigma_1(T_c)$ is much higher in magnitude and occurs
at higher values of the reduced temperature than in the model data.
It is clear from this analysis that the scattering rate obtained from
a TFM analysis cannot be used in BCS calculations to achieve a
quantitative understanding of the data. Nevertheless, it still has
some usefulness in that it allows one to understand qualitatively
how the collapse of the inelastic scattering at low temperatures can
result in a peak in the microwave conductivity at intermediate values
of the reduced temperature $t$.

In Figs.~\ref{fig:1} and \ref{fig:2} there is another set of open
(blue) down-triangles connected by a dotted line. These were obtained
in BCS calculations with $\tau^{-1}_\textrm{TFM-BCS}(T)/2$ as the
QP scattering rate in Eqs.~\eqref{eq:BCSa} and \eqref{eq:BCSb} but
now the gap is assumed to have $s$-wave symmetry $(x=1)$. We see that this
assumption has a drastic effect on both microwave conductivity and
penetration depth when compared with similar results for a $d$-wave
symmetric gap. In particular, the microwave conductivity does not show
a peak at intermediate reduced temperatures and is already very small
for $t=0.8$. This behavior is traced to the much more rapid drop in
normal fluid density with decreasing $T$ in the $s$ than in the
$d$-wave case. The first is exponentially small at $T\to 0$ while
the other is linear in $T$ in the same limit. This has important
implications for the data of Hashimoto {\it et al.}\cite{hashimoto09}
in  Ba$_{1-x}$K$_x$Fe$_2$As$_2$ as we will elaborate upon
in the next section.

\begin{figure}[tp]
  \vspace{-0.5cm}
  \includegraphics[width=9cm]{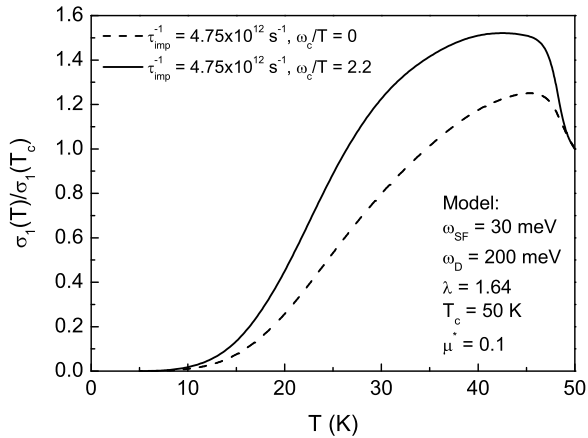}
  \caption{The normalized microwave conductivity $\sigma_1(T)/\sigma_1(T_c)$
at $28\,$GHz vs temperature $T$ for an isotropic $s$-wave Eliashberg
superconductor with the parameters noted in the figure and described in
the text. The solid and dashed lines include residual impurity
scattering $\tau^{-1}_{\textrm{imp}}$ of $4.75\times 10^{12}$
inverse seconds.
A low frequency cutoff in the $\alpha^2F(\omega)$ spectrum
of $\omega_c/T_c = 2.2$ was applied to simulate the collapse of the
inelastic scattering rate as $T\to 0$.
}
  \label{fig:4}
\end{figure}

Before turning to this discussion we make one final point about the pure
isotropic $s$-wave case. In Fig.~\ref{fig:4} we show results of
%
%
Eliashberg calculations with an MMP form for the electron-boson
interaction as we employed to describe the inelastic scattering in YBCO
but now an $s$-wave gap is used. The model parameters are $\omega_{SF} = 30\,$
meV, the  electron-spin fluctuation coupling strength $I^2$ was chosen
to give an mass enhancement factor of $\lambda = 1.64$, and the
high energy cutoff $\omega_D$ of the spectrum was set to $200\,$meV.
This resulted, together with the Coulomb pseudopotential
$\mu^\star = 0.1$ in a $T_c$ of $50\,$K more in line with the
critical temperatures observed in the ferropnictides.
Some elastic impurity scattering modeled by a constant value of the
impurity scattering rate $\tau^{-1}_{\textrm{imp}}=4.75\times 10^{12}\,$%
s$^{-1}$  was also included.
We note that the peak in the solid curve is at lower temperature
and is much broader than the usual coherence peak of BCS theory.
It also reacts to the addition of elastic impurity scattering
(dashed-dotted curve) in the opposite way to conventional BCS.\cite{mars91}
Increased impurity scattering depletes the magnitude of the peak, moves
it to higher temperatures and narrows it considerably. These effects
are the same as found in earlier work by  Nicol and Carbotte\cite{nicol91}
based on the MFLM of Varma {\it et al.}\cite{varma89} The physics
underlying the existence of the peak relates to scattering time
variations and not to the classical coherence argument of BCS theory.
What is important for the present paper is that the mechanism of the
collapse in inelastic scattering is much less effective in producing
peaks in the microwave conductivity for $s$-wave
than it is for $d$-wave gap symmetry. The fundamental difference
that accounts for this observation is that, in $s$-wave the normal
fluid density  drops to zero exponentially and becomes essentially
negligible at low temperatures where the $d$-wave normal fluid
density remains very significant.

We make a final point. While we discussed two different scattering
rates $\tau^{-1}_\textrm{TFM-BCS}(T)$ and $\tau^{-1}_\textrm{BCS}(T)$
one can define others which can be useful in different contexts.
For example, the
extended Drude model is often used to define a temperature and
frequency dependent optical scattering rate $\tau^{-1}_{op}(T,\omega)$
in terms of the complex optical conductivity, with
\begin{equation}
  \label{eq:Drude}
  \tau^{-1}_{op}(T,\omega) = \frac{\Omega^2_p}{4\pi}
  \textrm{Re}\left[\sigma^{-1}(T,\omega)\right].
\end{equation}
Evaluation of Eq.~\eqref{eq:Drude} for the microwave frequency
$\nu = 34.8\,$GHz gives the open (black) squares in Fig.~\ref{fig:3}
using our BCS model results. It is clear that
$\tau^{-1}(T,\nu=34.8\,\textrm{GHz})$ is totally different from either
$\tau^{-1}_\textrm{BCS}(T)$ or $\tau^{-1}_\textrm{TFM-BCS}(T)$. All
play a role depending on the question asked.\cite{schach03,hensen97,schach00}
Finally, we would like to note that just above $T_c$,
$\tau^{-1}_{op}(T=95\,\textrm{K})$ and
$\tau^{-1}_\textrm{TFM-BCS}(T=95\,\textrm{K})$ agree and are twice
the $\tau^{-1}_\textrm{BCS}(T=95\,\textrm{K})$.

\section{Two Band \mbox{\boldmath $s^\pm$} superconducting state}
\label{sec:3}

The newly discovered\cite{kamihara08} layered ferropnictide
superconductors display a complex band structure\cite{singh08} with
several electron and hole like pockets crossing the Fermi energy.
Multi-band superconductivity is now well established\cite{nicol05}
in MgB$_2$ which is widely believed to be a conventional
electron-phonon mechanism superconductor with two bands\cite{choi02,%
golubov02,jin03} one with a large gap and the other much smaller. While in the
ferropnictides there are more bands and the mechanism is not likely
to be the electron-phonon interaction,\cite{boeri08} a minimum model
that is often used is to include an electron band at the $M$ and a hole
band centered at the $\Gamma$ points of the Brillouin zone with
$s$-wave gaps of different magnitude with change in sign between the two
referred to as $s^\pm$-symmetry.\cite{mazin08} Angular resolved photo
emission spectroscopy (ARPES) in Ba$_{0.6}$K$_{0.4}$Fe$_2$As$_2$ by
Ding {\it et al.}\cite{ding08} gives values $\Delta_2\simeq 12\,$meV
and $\Delta_1\simeq 6\,$meV which were confirmed by Nakayama
{\it et al.}\cite{nakayama09} Somewhat smaller values $\sim 9\,$meV
and $\sim 4\,$meV are reported by the ARPES work of Evtushinsky
{\it et al.}\cite{evtushinsky09} An important question which remains
controversial is whether or not the $s$-wave gap on one of the Fermi
surfaces can be sufficiently anisotropic to acquire a node or is it
node-less.\cite{chubukov09,yanagi08,kuroki08,wang09,graser09}
Large anisotropies of the $s$-wave gap
are certainly expected even in conventional electron-phonon metals.
Indeed multiple plane wave calculations of the
electron-phonon spectral density $\alpha^2F(\omega)$ in Pb and
Al,\cite{tomlinson76,leung76a,leung76b} and first-principle calculations
of the electron-phonon contribution to the phonon linewidth
in Nb\cite{butler79} found large anisotropies
in this quantity and in the resulting superconducting gap values.
Also in the high $T_c$ cuprates where the gap has $d$-wave symmetry
calculations within a BCS spin fluctuation model with over-damped
magnons as in the work of Millis {\it et al.}\cite{millis90} have
produced gaps which go beyond the simplest $d$-wave versions
with many higher harmonics and even leads to
mixtures of $s$ and $d$-wave with profound effects on
the resulting temperature dependence of the penetration
depth.\cite{odon95a,odon95b,odon95c,branch95} The penetration depth
measurements in FeAs-122 of Hashimoto {\it et al.}\cite{hashimoto09}
gave isotropic gaps with $\Delta_2 \sim 6.8\,$meV and $\Delta_1 \sim
3.3\,$meV, considerably smaller than ARPES but, nevertheless, implying
an exponential activated behavior at low temperatures. In sharp contrast,
measurements by Martin {\it et al.}\cite{martin09} found a
non-exponential, close to $T^2$ law down to the lowest temperature
measured ($T\sim 0.2\,T_c$). While muon-spin resonance experiments
by Khasanov {\it et al.}\cite{khasanov09} gives isotropic gaps with
$\Delta_2\sim 9\,$meV and $\Delta_1\sim 1.5\,$meV we note
that this second gap is becoming rather small. Heat transport measurements
by Luo {\it et al.}\cite{luo09} are also consistent with no gap
nodes but they indicate that the anisotropic $s$-wave gap may be quite
small in certain momentum directions. Finally, we mention the nuclear
spin magnetic resonance data for $^{57}$Fe by Yashima {\it et al.}%
\cite{yashima09} which is consistent with $s^\pm$-symmetry with full
gaps.

\begin{figure}[tp]
  \vspace{-0.5cm}
  \includegraphics[width=9cm]{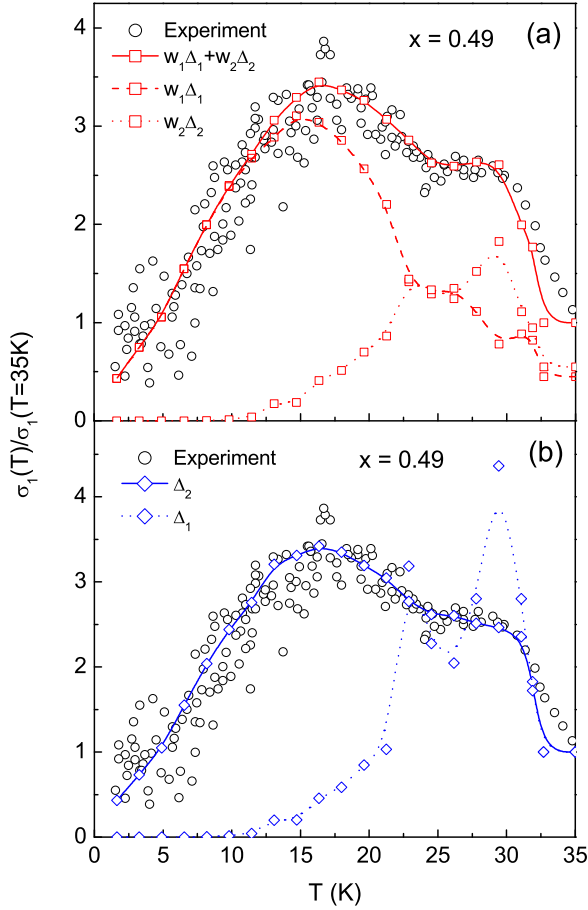}
  \caption{(Color online)
(a) Two band $s^\pm$ fit to the normalized microwave conductivity
$\sigma_1(T)/\sigma_1(T=35\,\textrm{K})$ of
Ref.~\onlinecite{hashimoto09} (open circles) at the microwave
frequency $\nu = 28\,$GHz vs temperature. In our model the small gap
$\Delta_1$ on the hole socket is assumed to be isotropic $s$-wave while
the large gap $\Delta_2$ on the electron pocket includes anisotropy
characterized by the parameter $x=0.49$ fixed to get a best fit to the
data [open (red) squares connected by a solid line]. The contribution
of the gap $w_1\Delta_1$ to the two band fit is shown by open (red) squares
connected by a dotted line while the open (red) squares connected by a
dashed line indicate the contribution of the second gap $w_2\Delta_2$.
The corresponding scattering rate $\tau^{-1}_\textrm{BCS}(T)$ is
shown by open (red) squares in the top frame of Fig.~\ref{fig:6}.
(b) The same as the top frame but now for a fit to experiment
using a single anisotropic $s$-wave gap $\Delta_2$ indicated by the open
(blue) chevron connected by a solid line. The anisotropy parameter is
$x=0.49$. 
}
  \label{fig:5}
\end{figure}

\begin{figure}[tp]
  \vspace{-0.5cm}
  \includegraphics[width=9cm]{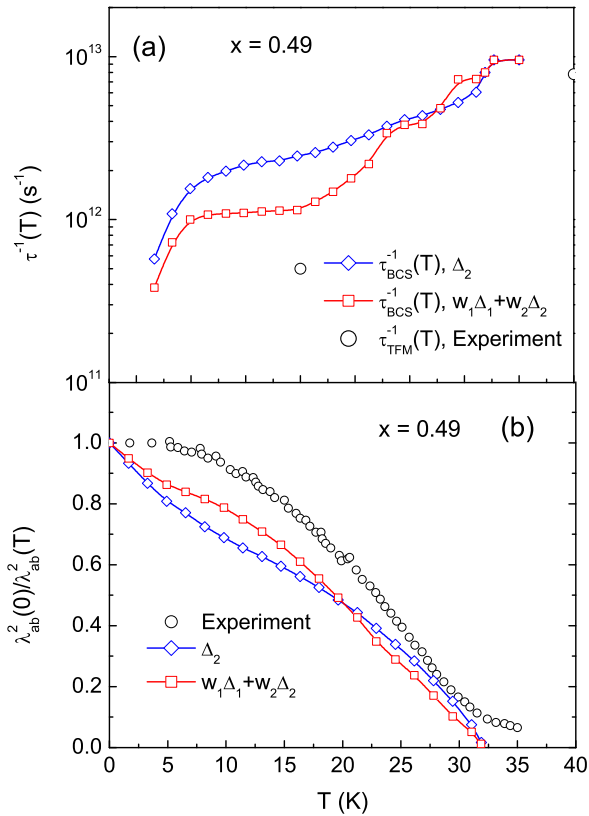}
  \caption{(Color online)
(a) The inelastic scattering rate $\tau^{-1}_\textrm{BCS}(T)$ in
inverse seconds obtained from our fits to the microwave conductivity
data (Ref.~\onlinecite{hashimoto09}) presented in
Fig.~\ref{fig:5}. The open (red) squares correspond to the fit
presented in the top frame of this figure while the open (blue)
chevron correspond to the fit presented in the bottom frame.
(b) The normalized inverse square of the penetration depth
$\lambda^2_{ab}(0)/\lambda^2_{ab}(T)$ vs temperature $T$. The open
(black) circles are the data for sample \#3 of Ref.~%
\onlinecite{hashimoto09}. The two curves indicated by open (red) squares
and open (blue) chevron, respectively have been generated using
BCS theory the temperature dependent inelastic scattering rates
indicated by the same symbols in the top frame of this figure.
The open circles are from the data of Table I of
Ref.~\onlinecite{hashimoto09}.
}
  \label{fig:6}
\end{figure}

In Fig.~\ref{fig:5}(a) we show results of our two band BCS calculations
%
%
based on Eqs.~\eqref{eq:BCS} for the normalized microwave conductivity
$\sigma_1(T)/\sigma_1(T=35\,\textrm{K})$ as a function of temperature $T$
for a microwave frequency $\nu = 28\,$GHz. The open (black) circles
indicate experimental results by Hashimoto {\it et al.}\cite{hashimoto09}
for FeAs-122, sample \#3 with $T_c = 32.7\,$K.
The open (red) squares connected by a solid line
show the theoretical results which are seen to fit well the data even at
low temperatures where the absorption appears to be roughly linear in $T$.
The corresponding temperature dependent inelastic
scattering rate $\tau^{-1}_\textrm{BCS}(T)$ is indicated by open (red)
squares in the top frame of Fig.~\ref{fig:6}. The small gap
$\Delta_1(0)$ on the hole surface was assumed to be isotropic $s$-wave
and equal to $3.3\,$meV [$\Delta_1(0)/(k_BT_c) = 1.17$] while the larger
gap of amplitude $\Delta_2(0) = 6.8\,$meV [$\Delta_2(0)/(k_BT_c) =
2.4$] was allowed to be anisotropic of the form
$\Delta_2(\theta) = \Delta_{sd}(\theta) =
\Delta_s+\Delta_d\sqrt{2}\cos(2\theta)$.
The contribution of the $s$-wave
gap to $\Delta_2$ is determined by the parameter $x$ of Eq.~\eqref{eq:x}.
The $s^\pm$-model is defined by the
combination of the two gaps $\Delta_{s^\pm} = w_1\Delta_1 + w_2\Delta_2$.
%
%
Furthermore, the complex conductivity $\sigma(T,\omega) =
w_1\sigma^{(1)}(T,\omega)+w_2\sigma^{(2)}(T,\omega)$. Here
$\sigma^{(1)}(T,\omega)$ and $\sigma^{(2)}(T,\omega)$ are the
complex conductivities calculated using Eq.~\eqref{eq:sigma} for
the two gapfunctions $\Delta_1(T,\omega)$ and $\Delta_2(T,\omega)$,
respectively. In doing so we are assuming that the interband transitions
can be ignored which is expected since the two $\sigma^{(i)}(T,\omega)$
originate from very different regions of momentum space.
In agreement with Hashimoto {\it et al.}\cite{hashimoto09} the weights
$w_1$ and $w_2$ have been chosen to be equal to 0.55 and 0.45, respectively.
Finally, the anisotropy parameter $x$ was allowed vary to get the best
fit to the microwave conductivity data over the whole temperature range
and was found to be equal to 0.49. Thus, we have a 49\%
$s$-wave contribution to
the big gap $\Delta_2$ which means that it has nodes
on the Fermi surface although it is anisotropic $s$-wave.\cite{schuerrer98}
The open (red) squares connected by a dotted line and a dashed line
give the individual contribution of $\Delta_1$ and $\Delta_2$, respectively,
to the microwave conductivity. It is clear from this decomposition that
it is the anisotropic gap with the nodes which contributes most at
low temperatures as well as to the peak at $\sim 17\,$K. However, the
hump around $30\,$K is due mainly, but not exclusively, to the small
gap. Note with reference to Fig.~\ref{fig:6}(a) that the temperature
dependent scattering rate obtained [open (red) squares] is very reasonable
and equal to $9.5\times 10^{12}\,$s$^{-1}$ at $T_c$ and shows a residual
scattering rate of $\tau^{-1}_\textrm{BCS}(T=1.6\,\textrm{K}) =
3.8\times 10^{11}\,$s$^{-1}$ indicating that sample \#3 of Hashimoto
{\it et al.}\cite{hashimoto09} is rather clean, so we do not expect
impurities to significantly alter the symmetry of the gap.
The open circles in
Fig.~\ref{fig:6}(a) are the optical scattering rates given in Table I
of Hashimoto {\it et al.}\cite{hashimoto09} obtained by a TFM analysis
of their microwave conductivity supplemented with their penetration
depth data on the same sample. As we found for the case of the cuprates
$\tau^{-1}_\textrm{TFM}(T)$ is quite different from
$\tau^{-1}_\textrm{BCS}(T)$ and is smaller by a significant amount. The
open (blue) chevrons are to be compared with the open (red) squares and
give the scattering rate according to BCS theory needed to fit the
microwave conductivity data in FeAs-122 with a single anisotropic gap
$\Delta_2$. The fit obtained is shown as the open (blue) chevrons
connected by a solid line in Fig.~\ref{fig:5}(b) which
nicely go through the experimental data (open circles). It is important
to note that if the the same scattering rate $\tau^{-1}_\textrm{BCS}(T)$
that fits experiment with the single anisotropic gap $\Delta_2$ is
used for the small isotropic $s$-wave gap $\Delta_1$ we get the open
(blue) chevrons with the dotted line through them. This curve is mainly
confined to the temperature region above $T=20\,$K and shows, once again,
that the symmetry of the gap has a determining affect on the temperature
variation of the resulting microwave conductivity.

Next we look at the temperature dependence of the normalized penetration
depth $\lambda^2_{ab}(0)/\lambda^2_{ab}(T)$ for the two gap $s^\pm$-model
which fits well the microwave conductivity of Fig.~\ref{fig:5}(a).
Theoretical results are shown as the open (red) squares on Fig.~\ref{fig:6}(b)
and are to be compared with the open circles which are the data for
sample \#3 of Hashimoto {\it et al.}\cite{hashimoto09} Once a
temperature dependent scattering rate $\tau^{-1}_\textrm{BCS}(T)$ was
fixed to fit the microwave conductivity there remained no adjustable
parameter and, as one can see, the fit to the data is deficient in two
ways. First of all, it is clear that the region near $T=0$ is not
exponentially activated as the data indicate but is
rather linear as one would expect from a gap with nodes. The data is
certainly more consistent with a node-less $s$-wave gap. Secondly, the
curve at higher temperatures falls quite a bit below the data.
If we had considered a single
anisotropic $s$-wave gap fit to the microwave conductivity data instead
of our two $s^\pm$-model we would have obtained the open (blue) chevrons
for the penetration depth which provide an even poorer over all fit. As
we described in the previous section on the cuprates we do not expect
even for a $d$-wave gap that we can fit equally well
the penetration depth with the
same scattering rate as determined from the fit to the microwave
conductivity, but for FeAs-122, the main problem has to do with the fact
that the anisotropy parameter $x=0.49$ obtained in the unconstrained
fit, leads to a large gap $\Delta_2$ which has nodes.

\begin{figure}[tp]
  \vspace{-0.5cm}
  \includegraphics[width=9cm]{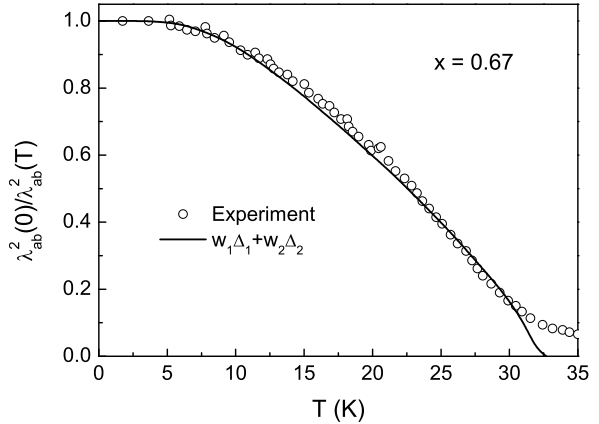}
  \caption{The normalized inverse square of the penetration depth
$\lambda^2_{ab}(0)/\lambda^2_{ab}(T)$ vs temperature $T$. The open
circles are the data for the FeAs-122 sample \#3 of Ref.~%
\onlinecite{hashimoto09}. The solid line represents the BCS fit to the
data using our two band $s^\pm$-model with isotropic $s$-wave smaller
gap $\Delta_1$ and an anisotropic larger gap $\Delta_2$ with the
anisotropy parameter $x=0.67$ in the clean limit. The weights
$w_1 = 0.55$ and $w_2 = 0.45$.
}
  \label{fig:7}
\end{figure}

A natural question to ask next is what would happen if instead of fitting
the microwave conductivity we fit the penetration depth. In Fig.~\ref{fig:7}
%
%
we consider the BCS clean limit and vary the anisotropy parameter $x$
to get a good fit to the data (open circles). For $x = 0.67$ we get the
solid line which is in very good agreement with experiment. It corresponds
to the mixture $w_1\Delta_1+w_2\Delta_2$ of the two gaps in our
$s^\pm$-model with $w_1 = 0.55$ and $w_2 = 0.45$. We proceed with these
new $s^\pm$-model parameters to find a temperature dependent scattering
rate to fit the microwave conductivity data. Our results are shown as
%
%
\begin{figure}[tp]
  \vspace{-0.5cm}
  \includegraphics[width=9cm]{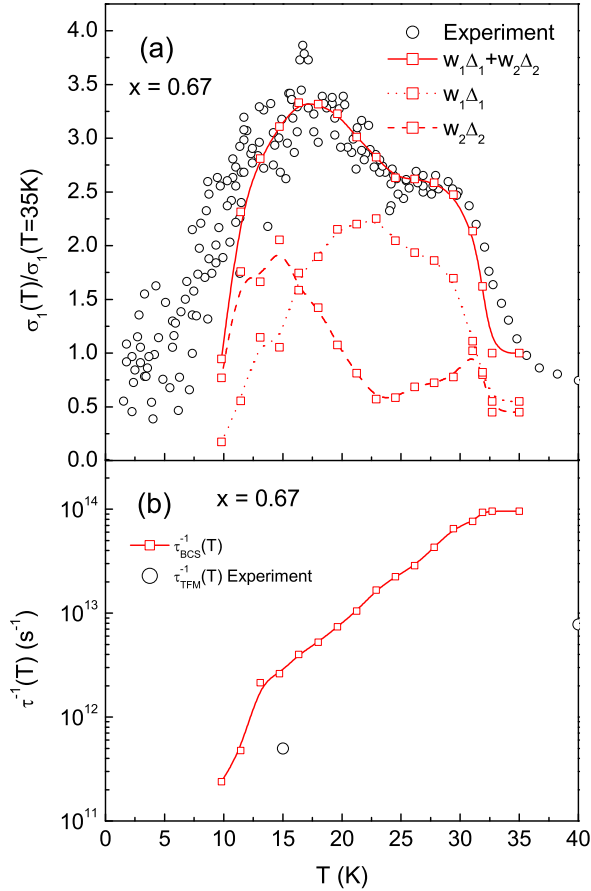}
  \caption{(Color online)
(a) The normalized microwave conductivity $\sigma_1(T)/
\sigma_1(T=35\,\textrm{K})$ for the FeAs-122 sample \#3 of
Ref.~\onlinecite{hashimoto09} at $\nu = 28\,$GHz as a
function of temperature $T$. The open (red) squares connected by a
solid line are the best fit obtained within a two band $s^\pm$-band
BCS theory with a temperature dependent scattering rate
$\tau^{-1}_\textrm{BCS}(T)$ as to simulate the inelastic scattering.
The smaller (hole band) gap is isotropic $s$-wave and the larger
(electron band) gap is anisotropic $s$-wave with $x=0.67$ chosen to
give a reasonable fit to the penetration depth data. (See Fig.~\ref{fig:7}.)
The open (red) squares connected to the dashed and dotted lines give
the individual contributions of the two gaps.
(b) The inelastic scattering rate $\tau^{-1}(T)$ vs temperature. The
open (red) squares give $\tau^{-1}_\textrm{BCS}(T)$, the inelastic
scattering rate from our fit to the microwave conductivity shown in
the top frame of this figure. The open circles represent data
for the FeAs-122 sample \#3 of Ref.~\onlinecite{hashimoto09}.
}
  \label{fig:8}
\end{figure}
the solid (red) squares connected by a solid line in Fig.~\ref{fig:8}(a).
The over all fit is good including the region of the peak at $\sim 17\,$K.
The main deficiency is that below $10\,$K where the theoretical curve drops
sharply to zero while the experiment still gives absorption. There
are two features of this fit we wish to emphasize. The open (red) squares
connected by a dotted line give the separate contribution to the total
from the small gap $w_1\Delta_1$ while the dashed curve with the open
(red) squares is from the large, now node-less, anisotropic gap $w_2\Delta_2$.
In contrast to what was observed in Fig.~\ref{fig:5}(a) where the large
gap dominated the low temperature behavior and the small gap contributed
mainly just below $T_c$ now both contributions extend over all temperatures
with the small gap contribution still important at lowest temperatures and
displaying a broad peak at $\sim 23\,$K.
The BCS inelastic scattering rate
obtained from this fit is given in Fig.~\ref{fig:8}(b) by open (red)
squares. We note in comparison with the data of Fig.~\ref{fig:6}(a)
that $\tau^{-1}_\textrm{BCS}(T=T_c)$ is now much bigger and approximately
equal to $1.9\times 10^{14}\,$s$^{-1}$. This is much bigger than the
scattering rate derived from a TFM fit to the data by Hashimoto {\it
et al.}\cite{hashimoto09} shown as the open circles. While
$\tau^{-1}_\textrm{BCS}(T=T_c)$ is to be interpreted as due to inelastic
scattering it is rather big and indicates that this second
fit to the microwave conductivity data while more compatible with the
experimental penetration depth at low temperatures
remains somewhat problematical.

\begin{figure}[tp]
  \vspace{-0.5cm}
  \includegraphics[width=9cm]{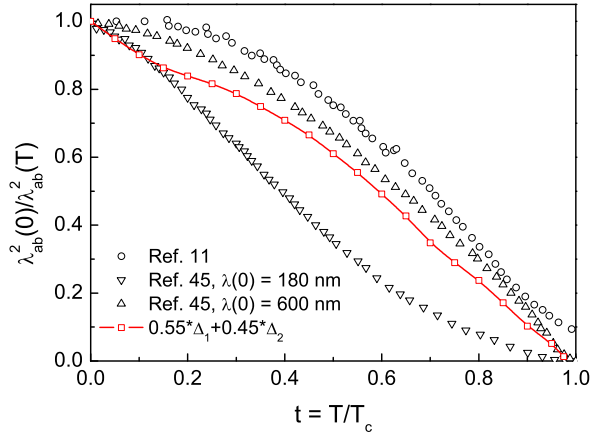}
  \caption{(Color online) The normalized inverse square of the
penetration depth $\lambda^2_{ab}(0)/\lambda^2_{ab}(T)$ vs the reduced
temperature $t$. The open circles represent the data by Hashimoto
{\it et al.}\cite{hashimoto09} while the open down-triangles and
open up-triangles give the results reported by Martin {\it et al.}%
\cite{martin09} for $\lambda_{ab}(0)$ equal to $180\,$nm and $600\,$nm,
respectively. The open (red) squares correspond to our result of
a BCS calculation in $s^\pm$ symmetry where the large gap $\Delta_2$
has nodes on the electronic Fermi surface [see also Fig.~\ref{fig:6}(b)].
}
  \label{fig:9}
\end{figure}

Figure~\ref{fig:9} presents a comparison between experimental results
%
%
of Hashimoto {\it et al.}\cite{hashimoto09} (open circles) and of
Martin {\it et al.}\cite{martin09} [open down-triangles for $\lambda_{ab}(0)
= 180\,$nm and open up-triangles for $\lambda_{ab}(0) = 600\,$nm].
Our result of a BCS $s^\pm$-symmetry calculation is indicated by the
open (red) squares. They correspond to an anisotropic large gap
$\Delta_2$ with nodes on the electronic Fermi surface and have already
been discussed in Fig.~\ref{fig:6}(b). At low temperatures theory
agrees well with the data by Martin {\it et al.}\cite{martin09} for
$\lambda_{ab}(0) = 180\,$nm. For higher temperatures theory is always
above this data set but stays significantly below the data for
$\lambda_{ab}(0) = 600\,$nm. As $\lambda_{ab}(0)$ plays the role of
a fitting parameter in the analysis of Martin {\it et al.}\cite{martin09}
a $\lambda_{ab} \simeq 400\,$nm will probably bring experiment closer
to theory. Nevertheless, the low temperature dependence of the
penetration depth as reported by Martin {\it et al.}\cite{martin09}
is certainly more in line with the microwave conductivity data
of Hashimoto {\it et al.}\cite{hashimoto09} which shows a linear
temperature dependence of $\sigma_1(T)/\sigma_1(T=35\,\textrm{K})$
below $T=10\,$K for $T\to 0$. Thus, at low temperatures there is
still substantial absorption in the system which is in contradiction to
the exponentially activated behavior observed by Hashimoto {\it et al.}%
\cite{hashimoto09} In a final point it is certainly important to note
that the potassium content is quite different in the samples used
in both experiments. Hashimoto {\it et al.}\cite{hashimoto09} report
for their sample \#3 $\sim55\,$\% potassium while Martin {\it et al.}%
\cite{martin09} report a potassium content of $\sim 30\,$\% for their
sample B.

\section{Summary and conclusion}
\label{sec:4}

We have reexamined the use of the two fluid model as a way of extracting
from a combination of microwave conductivity and penetration depth data
a temperature dependent scattering rate
$\tau^{-1}_\textrm{TFM}(T)$ which is believed to model the inelastic
scattering. It has the very desirable property that, when it is
multiplied into the normal fluid density, it reproduces the microwave
conductivity. However, no test of its general validity has been provided.
In this paper we take a different approach and use instead BCS theory
to extract through a tight fit to the microwave conductivity a
new temperature dependent inelastic scattering rate
$\tau^{-1}_\textrm{BCS}(T)$. A comparison of $\tau^{-1}_\textrm{BCS}(T)$
with its TFM counterpart shows that they differ significantly both
in absolute magnitude and in variation with $T$. This casts doubts
on the quantitative significance of $\tau^{-1}_\textrm{TFM}(T)$.
Nevertheless, the TFM does indeed provide a very useful basis for a
first understanding of the role of inelastic scattering in those
phenomena.

The fact that penetration depth information was not needed to extract our
$\tau^{-1}_\textrm{BCS}(T)$
provides a first test of its validity. We use it in a
BCS calculation with $d$-wave gap symmetry and find good semiquantitative
agreement with the data in optimally doped YBCO, the same material
used for the fit to the microwave conductivity. We take this small, yet
significant discrepancy, as evidence that a temperature dependent but
constant in frequency QP scattering rate does not capture all of the
quantitative features of inelastic scattering. Because of the
simplifications inherent in BCS theory a constant in frequency scattering
rate is strictly required. But inelastic scattering intrinsically
implies a frequency dependence to $\tau^{-1}(T,\omega)$ which is closely
linked to its $T$ dependence. For instance for electron-phonon coupling
the $\omega^2$ dependence of the spectral density $\alpha^2F(\omega)$
at small $\omega$ implies a $T^3$ law for the quasiparticle scattering
rate at $\omega=0$ and also a $\omega^3$ law for $T=0$. Similarly, for
the coupling to over-damped spin fluctuations as is the case in the
nearly antiferromagnetic Fermi-liquid model of the cuprates the
corresponding laws are $T^2$ and $\omega^2$. Furthermore, the frequency
dependence of $\tau^{-1}(T,\omega)$ implies by Kramers-Kronig transform
a nontrivial (i.e.: nonzero) real part of the quasiparticle self
energy. It is precisely these features that are incorporated in the
$d$-wave generalization of Eliashberg theory which we have used in
our previous work. It is the results of our previous quantitative fit to
microwave conductivity and penetration depth data in optimally doped
YBCO single crystals that we have used here in our comparison with BCS.
An important conclusion of all this is that Eliashberg theory is fully
quantitative while BCS provides a good semiquantitative picture, its
main limitation being due to the neglect of frequency dependence of
the inelastic QP scattering rate.

Having established the validity of BCS theory to provide meaningful
results in the present context we considered next its generalization
to the two band case. This is the minimum model required for a realistic
treatment of superconductivity in the ferropnictides in which there
can be several hole and electron pockets with different values of the
superconducting gap. A favored model is an $s^\pm$-model with
isotropic $s$-wave gaps on each of the two bands and with opposite signs.
We also consider the possibility that one of the gaps is anisotropic
$s$-wave as in the work of Chubukov {\it et al.}\cite{chubukov09}
and others.\cite{wang09} Anisotropy is expected
even from consideration of conventional superconductivity\cite{graser09,
tomlinson76,leung76a,leung76b,butler79} and also the
cuprates.\cite{odon95a,odon95b,odon95c}

A general conclusion of our analysis is that with isotropic $s$-wave
it is difficult to get realistic values of the quasiparticle scattering
rate $\tau^{-1}_\textrm{BCS}(T)$ from microwave conductivity data
which shows very significant absorption at low temperatures as is seen
in the data of Hashimoto {\it et al.}\cite{hashimoto09} for FeAs-122.
This is also the case when the big gap is allowed to be anisotropic
but with no nodes on the Fermi surface. On the other hand, if the
gap is sufficiently anisotropic to have nodes it becomes rather easy
to get a fit to the FeAs-122 microwave conductivity data with realistic
values of $\tau^{-1}_\textrm{BCS}(T)$. But a node on one of the gaps
also leads to a penetration depth which shows $d$-wave like low
temperature power laws (i.e.: linear in temperature) rather than the
exponential activation behavior found experimentally on the same
FeAs-122 sample which is the characteristic of a gap with finite value
everywhere on the Fermi surface. On the other hand, the microwave
conductivity data of Hashimoto {\it et al.}\cite{hashimoto09} is
much more consistent with the more recent penetration depth data
of Martin {\it et al.}\cite{martin09} on a related but not identical
FeAs-122 sample.

\acknowledgements
Research supported in part by the Natural Sciences and Engineering Research
Council of Canada (NSERC) and by the Canadian Institute for Advanced Research
(CIFAR).
%
%
\bibliography{feas2}
\end{document}